\documentclass[twocolumn,showpacs,eqsecnum]{revtex4}
\usepackage{xspace,epsfig,psfig,amsfonts,amssymb}

\begin{document}

\newcommand{\beq}{\begin{equation}}
\newcommand{\eeq}{\end{equation}}
\newcommand{\bqa}{\begin{eqnarray}}
\newcommand{\eqa}{\end{eqnarray}}
\newcommand{\nn}{\nonumber}
\newcommand{\nl}[1]{\nn \\ && {#1}\,}
\newcommand{\erf}[1]{Eq.~(\ref{#1})}
\newcommand{\dg}{^\dagger}
\newcommand{\rt}[1]{\sqrt{#1}\,}
\newcommand{\smallfrac}[2]{\mbox{$\frac{#1}{#2}$}}
\newcommand{\half}{\smallfrac{1}{2}}
\newcommand{\bra}[1]{\langle{#1}|}
\newcommand{\ket}[1]{|{#1}\rangle}
\newcommand{\ip}[2]{\langle{#1}|{#2}\rangle}
\newcommand{\sch}{Schr\"odinger }
\newcommand{\schs}{Schr\"odinger's }
\newcommand{\hei}{Heisenberg }
\newcommand{\heis}{Heisenberg's }
\newcommand{\bl}{{\bigl(}}
\newcommand{\br}{{\bigr)}}
\newcommand{\ito}{It\^o }
\newcommand{\str}{Stratonovich }
\newcommand{\dbd}[1]{{\partial}/{\partial {#1}}}
\newcommand{\sq}[1]{\left[ {#1} \right]}
\newcommand{\cu}[1]{\left\{ {#1} \right\}}
\newcommand{\ro}[1]{\left( {#1} \right)}
\newcommand{\an}[1]{\left\langle{#1}\right\rangle}
\newcommand{\implies}{\Longrightarrow}
\newcommand{\tr}[1]{{\rm Tr}\sq{ {#1} }}
\newcommand{\del}{\nabla}
\newcommand{\du}{\partial}
\newcommand{\nb}{\bar{n}}

\newcommand{\rb}[1]{\raisebox{1.5ex}[0pt]{#1}}

\title{Adaptive quantum-limited estimates of phase}

\author{Written by: H.\ M.\ Wiseman}
\affiliation{Centre for Quantum Dynamics, School of Science, Griffith 
University, Nathan 4111, Australia}
\author{Latest Theory by: D.\ W.\ Berry}
\affiliation{Department of Physics and Centre for Advanced Computing 
-- Algorithms and Cryptography, Macquarie University, Sydney 2109, 
Australia}
\author{Experiments by: H. Mabuchi and co-workers}
\affiliation{Norman Bridge Laboratory of Physics 12-33, California
Institute of Technology, Pasadena, CA 91125, USA}

\begin{abstract}
Quantum-limited estimation of an optical phase using 
adaptive (i.e. real-time feedback) techniques is reviewed. One case is 
explored in detail, as it can be understood using only elementary 
concepts such as photonic shot-noise and error analysis. 
Very recent experimental 
results are discussed.
\end{abstract}
\maketitle

\section{Introduction}
Having been working on the topic of quantum-limit adaptive estimation of 
optical phase for some years now 
\cite{Wis95,WisKil97,WisKil98,BerWisZha99,BerWis00,BerWis01a,%
BerWis01b,BerWisBre01,BerWis02}, when I (HMW) was asked to contribute an 
article on it, the question ``why now?'' naturally arose. I believe 
there are three good 
answers to this question. In order of increasing importance,
\begin{enumerate}
\item The work done by me, and more particularly my students, covers a 
wide range of cases. It is only now that it is possible to put them 
all in an overall context.
\item Some of the latest results are actually the easiest to 
explain to a non-specialist.
\item The first experiments verifying the theory have very 
recently been performed in the laboratory of Hideo Mabuchi at CalTech.
\end{enumerate}

Before discussing these interesting developments, I should explain 
what quantum-limited adaptive phase estimation means. The phase in 
question is an  unknown (and possibly varying) optical phase 
${\varphi}$. The {\em aim} is to estimate this phase, on the basis of 
measurements, as well as quantum mechanics allows. The {\em allowed resources} 
are practical ones: photodetectors, electronics, and linear 
(electro-)optics. The {\em fundamental limitation} is the 
number of photons (per pulse or per 
coherence time). There are also many {\em practical limitations} such 
as efficiency, the source of light, time 
delays, and processing limitations. These have all been considered 
\cite{WisKil97,BerWisZha99,BerWis01a,BerWis01b}, but for simplicity I will not 
discuss them here.

Where {\em adaptive} estimation comes in is that to realise the above 
aim using the allowed resources 
it turns out that it is necessary to use {\em real-time} feedback 
during the measurement. The basic idea is as follows. 
If ${\varphi}$ is known to a very good approximation, then 
a simple measurement 
scheme will usually give near-optimal results. An example is 
homodyne measurement of the phase quadrature of a pulse 
\cite{WalMil94}.
However if ${\varphi}$ is totally unknown then the standard solution is to 
measure both quadratures (e.g. by heterodyne detection). This is far from 
optimal. The alternative is adaptive detection:  
use the results {\em so far} in the measurement 
to make an {estimate} ${\hat\varphi}$ of ${\varphi}$, and 
use this in a {feedback loop} to make 
the rest of the measurement closer to optimal.

I believe this work is important for a number of reasons:

First, 
quantum phase has a long and controversial history \cite{PegBar97}. 
Although ideal phase measurements can be defined, there is no way to
make them without optical materials with arbitrarily high orders of 
nonlinearity \cite{WisKil98}. Hence it is of fundamental interest to
know how well one can do with linear devices. 

Second, with continued miniaturisation 
of devices there will come a time when quantum limits do 
set the fundamental limits for technology. There is every reason to 
expect that phase estimation, as required for phase locking loops for 
example, will still be a practical concern then. 

Third, the idea of adaptive measurements may have applications far 
beyond that of phase measurements, so the theoretical and 
experimental expertise gained here potentially opens new frontiers in 
quantum measurement theory. It also has implications for the way 
in which we understand phenomena such as the ``collapse of the 
wavefunction''.

Finally, the cutting-edge technology required for 
quantum-limited adaptive measurements is a stepping stone 
towards more general methods for controlling quantum systems, 
another area of future importance as engineering 
heads towards the quantum realm. As will be discussed in Sec.~V, 
the new technology being applied is both electronic and optical. 

\section{Overview of Past Results}

There are many different cases that can be considered, and they can be 
systematised by considering the following four questions:
\begin{enumerate}
\item Is the detection {\em dyne} (that is, using an optical local oscillator) 
or {\em interferometric}?

\item Is the light source {\em coherent} (e.g. a laser), or {\em nonclassical} 
(e.g. squeezed)?

\item Is the scheme {\em non-adaptive}, or {\em adaptive}?

\item Is it {\em single-shot} with a single phase, or {\em CW} 
with a varying phase? 
\end{enumerate}
Two of these distinctions definitely require further explanation. 

First, {\em dyne versus interferometric} detection. These are illustrated in 
the figures below. 

\begin{figure}[h] 
\includegraphics[width=0.4\textwidth]{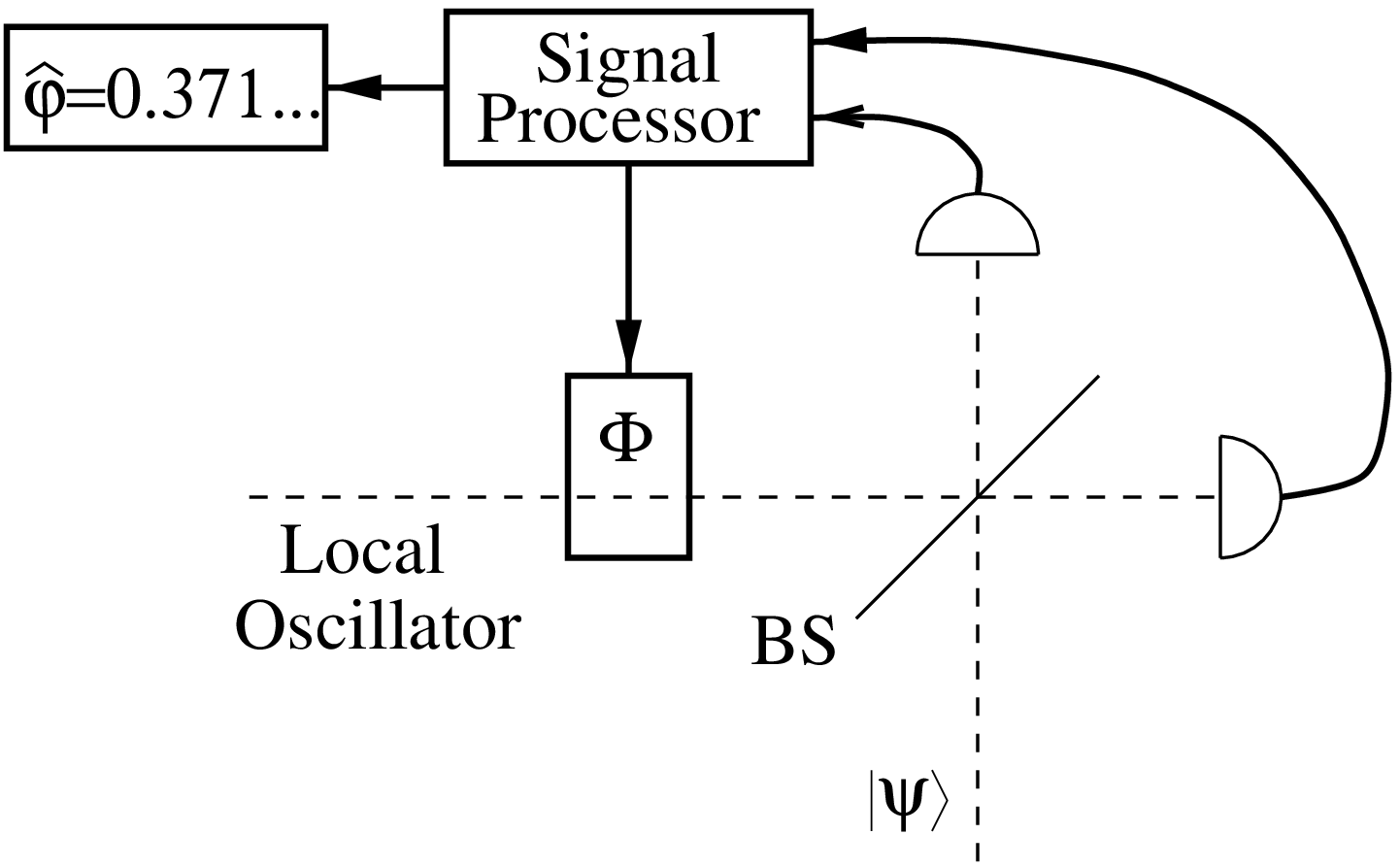} 
\end{figure}

\begin{figure}[h]
\includegraphics[width=0.4\textwidth]{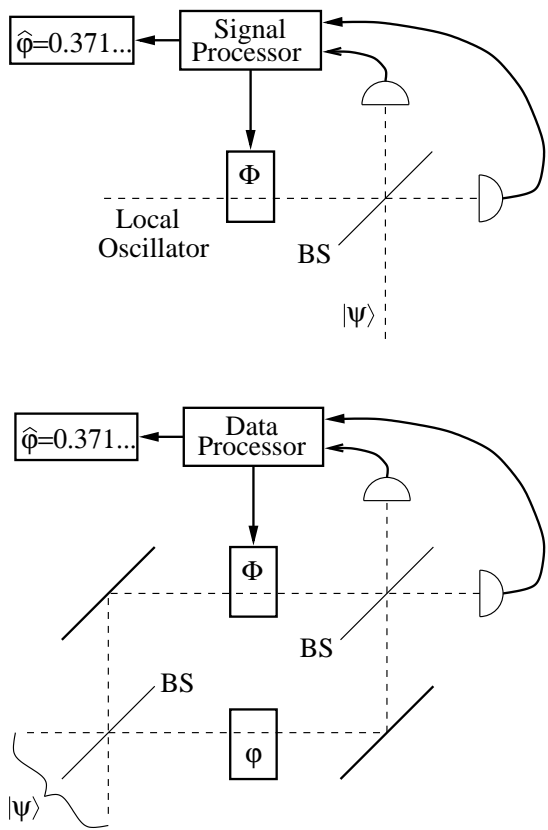}
\caption{Adaptive phase estimation: dyne detection (top) and 
interferometric detection (bottom). BS = Beam Splitter, and dashed lines 
are optical paths.}
\end{figure}

It might well be 
objected that dyne detection (of which homodyne and heterodyne are two 
examples \cite{WisKil97}) is merely one sort of interferometric 
measurement \cite{HarSan96}, in that the local oscillator is interfered with the 
signal light before detection. This is quite true; I am using the two
terms to make a distinction between how the accounting is done. In 
dyne detection only the signal pulse or beam is treated as quantum, 
and only the number of photons in it is counted for the 
purposes of determining the quantum limit. In interferometric detection
both input pulses or beams into the Mach-Zehnder interferometer are 
treated as quantum, and the photon number in both is counted 
\cite{BerWis00,BerWisBre01}. There is 
also a practical distinction in that dyne detection uses a 
photoreceiver that yields a photocurrent which does not distinguish 
individual photons, whereas interferometric detection uses photon 
counters. In both cases the aim is to measure the phase $\varphi$; in 
the former this phase is defined  relative to the 
 local oscillator, while for the latter it is as a relative phase between 
 the two arms of the interferometer.

Second, {\em single-shot versus CW}. In a single-shot measurement, 
there is a single pulse of light (which may however be split over two 
modes for the case of interferometric detection). There is a single 
 unknown phase $\varphi$ imprinted on the pulse.  The relevant parameter 
 for the fundamental limit is $n$ (or $\bar{n}$), 
 the (mean) number of photons in the pulse. In {\em Continuous-Wave} 
 detection, there is a beam (or two beams in the interferometric case) 
 of light. A time-varying phase $\varphi(t)$ is 
 imprinted on (one of) the beam(s). In this article I will consider 
 only the case where the time variation is that of white noise. That 
 is, 
 \beq
 \dot{\varphi} = \sqrt{\kappa}\,{\xi}(t).
\eeq
This could arise from thermal mechanical fluctuations 
of miniaturised optical elements, for example. The parameter $\kappa$ is the rate 
of phase diffusion, or, equivalently, the resulting linewidth of the 
beam. The relevant parameter in this case 
is $N$, the mean number of photons 
per coherence time: $N = P/\hbar\omega{\kappa}$, where $P$ is the beam 
power.

Having explained the various cases, I can now give the promised 
overview. The simplest results arise by considering 
the mean square error (MSE) in the estimated phase, in the asymptotic limit 
where $n$ or $\bar{n}$ or $N$ go to $\infty$. The results for the 
MSE scale as in table 1.

\begin{table}[h]
\begin{tabular}{|c|c|c|c|c|c|}
    \cline{3-6}
    \multicolumn{2}{c|}{$\phantom{\stackrel{?}{\sim}}$} & \multicolumn{2}{c|}{dyne} & 
    \multicolumn{2}{c|}{interferometric} \\
	\cline{3-6}
	\multicolumn{2}{c|}{$\phantom{\stackrel{?}{\sim}}$} & coherent & 
	{non-class.} & coherent  &{non-class.}  \\
	\hline
	  & {adaptive}  & {$0.5/N^{1/2}$} &  $ \underline{\sim 
	  1/N^{2/3}}$ & {$1/N^{1/2}$} & 
	${?}\phantom{\stackrel{?}{\sim}}$ \\
	\cline{2-6}
	\rb{CW} & non-adapt. & $0.71/N^{1/2}$ & $ {0.66/N^{1/2}}$ & $1/N^{1/2}$ & 
	${?}\phantom{\stackrel{?}{\sim}}$\\
	\hline
	single & {adaptive}  & {$0.25/\nb$} & 
	$ \underline{\sim \log \nb / \nb^{2}}$ & 
	{$1/n$} &  $\underline{\stackrel{?}{\sim} \log n / n^{2}}$ \\
	\cline{2-6}
	shot & non-adapt. & $0.5/\nb  $ & ${0.25/\nb} $ & $1/n$ &
	${\sim 1/n} \phantom{\stackrel{?}{\sim}}$ \\
	\hline
\end{tabular}
\caption{Asymptotic mean square errors for phase estimation. The 
results that beat the standard quantum limits are underlined.}
\end{table}

There is a good deal of regularity in this table, as the reader may 
discern. I wish to draw attention to one feature in particular. The 
sixteen MSEs (some still undetermined) can be divided into 
four squares of four MSEs. In each square for which results are fully known, 
three of the MSEs scale in the same way. 
This scaling represents the {\em standard 
quantum limit} (SQL) for that particular case. The fourth, underlined 
in the above 
table, beats the SQL. In each case, this requires 
both {\em non-classical light} and {\em adaptive detection}.

\section{Latest Results: CW Dyne Detection}

To try to explain adaptive phase estimation in more detail, I will 
concentrate now on the particular case of CW dyne detection. This is 
one the latest areas to be investigated, by Dominic Berry and myself 
\cite{BerWis02}, but turns out to be probably the simplest to explain. 
To reiterate the basic idea, we want the 
 best estimate of the {\em current value} of $\varphi(t)$ which obeys
$ \dot{\varphi} = \sqrt{\kappa}\,{\xi}(t),$
where $\xi(t)$ is white noise with unit spectral power.

Consider first the case of {\em coherent light} of power 
${P}=\hbar\omega{\alpha}^{2}$, detected by interfering with a local 
oscillator at a balanced photoreceiver. The resulting dyne 
photocurrent, suitably scaled, is \cite{WisKil97}
\beq
{I}(t) = 2{\alpha} \cos[\Phi(t)-{\varphi}(t)] + 
{\zeta}(t), \label{dynepc}
\eeq
where $\Phi(t)$ is the local oscillator phase, and $\zeta(t)$ is 
another Gaussian white noise term, independent of $\xi(t)$, with unit 
spectral power.
It can be thought of as local oscillator shot noise, or vacuum 
noise. In any case, it is {\em quantum noise}.

The standard non-adaptive technique to estimate phase 
is to vary $\Phi(t)$ over all phases. This can be achieved by 
 heterodyne detection with a detuning $\Delta \gg 
 \sqrt{\alpha\kappa}$, which makes $\Phi$ vary as 
$\Phi(t) = \Phi(0) + \Delta\times t$.
From \erf{dynepc}, we would 
expect better sensitivity if we were to choose $\Phi(t)$  to maximise the 
slope of $\cos[\Phi(t)-{\varphi}(t)]$. 
That is, we should set 
${\Phi}(t) = {\hat\varphi}(t) + \pi/2$ so that
\bqa
{I}(t) &=& 2{\alpha} \sin[{\varphi}(t) - {\hat\varphi}(t)] 
+ {\zeta}(t) \nn \\
&\simeq& 2{\alpha}[{\varphi}(t) - {
\hat\varphi}(t)]+{\zeta}(t), \label{simeq1}
\eqa
for ${\hat\varphi}(t) \simeq {\varphi}(t)$. 
The question is, how should we choose the estimate ${\hat\varphi}(t)$? 

One obvious possibility is to choose it from the immediately preceding
photocurrent. We can rearrange \erf{simeq1} to get 
\beq
{\varphi}(t) \simeq \sq{{\hat\varphi}(t) + \frac{{
I}(t)}{2{\alpha}}} + 
\frac{{\zeta}(t)}{2{\alpha}}.
\eeq
For a given ${\hat\varphi}(t)$, we could thus form 
\beq
{\hat\varphi_{\rm imm}}(t+\delta t) = 
\frac{1}{\delta t}\int_{t}^{t+\delta t}
\sq{{\hat\varphi}(t) + \frac{{I}(s)}{2{\alpha}}}ds.
\eeq
This has the MSE 
\bqa
{\sigma^{2}_{\rm imm}}(t+\delta t) &=& \an{[{\hat\varphi_{\rm 
imm}}(t+\delta t)-{\varphi}(t+\delta t)]^{2}} \nn \\
&\simeq&  
\frac{1}{4{\alpha}^{2}\delta t}.
\eqa
This diverges as $\delta t \to dt$, so clearly ${\hat\varphi_{\rm 
imm}}$ is not a good estimate. Instead, we need a ${\hat\varphi}(t)$ 
that involves a finite time 
average.

The optimal time-average for ${\hat\varphi(t)}$ can be determined as 
follows. 
Say the MSE in ${\hat\varphi}(t)$ is ${\sigma^{2}}(t)$. Over the interval 
$[t,t+\delta t)$, the diffusion of ${\varphi}(t)$ causes this to increase to
\bqa 
{\sigma^{2}_{\rm old}}(t+\delta t) &=& \an{[{\hat\varphi}(t)-{
\varphi}(t+\delta t)]^{2}} 
\nn \\ &=& 
\an{[{\hat\varphi}(t)-{\varphi}(t)]^{2}}+
 \an{[{\varphi}(t)-{\varphi}(t+\delta t)]^{2}}  \nn \\
&=& 
{\sigma^{2}}(t) + {\kappa} \delta t.
\eqa
By standard error analysis, the optimal ${\hat\varphi}(t+\delta t)$ weights ${
\hat\varphi}(t)$ and 
${\hat\varphi_{\rm imm}}(t+\delta t)$ appropriately:
\beq {\hat\varphi}(t+\delta t) = {\sigma^{2}}(t+\delta 
t)\sq{\frac{{\hat\varphi_{\rm imm}}(t+\delta t)}{{\sigma^{2}_{\rm 
 imm}}(t+\delta t)} + \frac{{\hat\varphi}(t)}{{\sigma^{2}_{\rm  old}}
 (t+\delta t)}}, \label{forhatvp}
 \eeq
where
\beq
\frac{1}{{\sigma^{2}}(t+\delta t)} = \frac{1}{{\sigma^{2}_{\rm 
 imm}}(t+\delta t)} + \frac{1}{{\sigma^{2}_{\rm  old}}(t+\delta t)}.
 \eeq
 
Taking $\delta t \to dt$ yields a differential equation for $\sigma^{2}(t)$ 
that has the stationary solution
\beq \sigma^{2} = 1/2\sqrt{{N}}~,~~{\rm where}~{N}={
\alpha}^{2}/{\kappa} . \label{adaptSQL}
\eeq
Substituting this into \erf{forhatvp} gives
$ d{\hat\varphi}(t) = (\kappa/\sigma^{2})\times({I}(t)dt/2{\alpha}).$ 
Since the feedback sets $\Phi(t) = \hat\varphi(t)+\pi/2$, the {\em feedback 
algorithm} is simply
\beq
\dot{\Phi} = \frac{\kappa}{\sigma^{2}}\frac{{I}(t)}{2{\alpha}}.
\eeq

By way of comparison, for heterodyne (nonadaptive) detection, the stationary MSE can be 
shown to be \cite{BerWis02}
\beq \label{hetlim}
\sigma^{2}_{\rm het} = 1/\sqrt{2{N}}.
\eeq
That is, the adaptive technique offers a factor of $1/\sqrt{2}$ 
improvement in the mean square error.

\section{What is quantum about it?}

This is a question I am often asked. 
A feedback algorithm of the form
\beq \label{genfb}
\dot{\Phi} = {\chi}{I}(t)/2{\alpha}
\eeq 
looks like a standard low frequency classical 
{phase-locking algorithm} with gain $\chi$.
The quantum feature is that the optimal value of ${\chi}$ is finite:
\beq
{\chi_{\rm opt}}  = \frac{\kappa}{\sigma^{2}} = 2\sqrt{
\kappa}{\alpha}. 
\eeq
In the classical limit (with no other noise sources), 
$\chi_{\rm opt}$ would be infinite, as can be 
seen by writing it explicitly in terms of classical quantities and 
$\hbar$:
\beq
\chi_{\rm opt} =  
2\sqrt{{\kappa}{P} /\hbar\omega}.
\eeq

For general ${\chi}$, the stationary MSE can be shown to be
\beq
\sigma^{2} = \frac{\chi}{8{\alpha}^{2}}+ \frac{\kappa}{2{\chi}},
\eeq
which diverges as ${\chi} \to \infty$. Moreover, if $\chi$ differed
from $\chi_{\rm opt}$ by a factor greater than about $2.4$ (in either 
direction), then all advantage gained by doing an adaptive detection 
would be lost. That is, the MSE would become greater than that which 
can be achieved by the nonadaptive technique of heterodyne 
detection (\ref{hetlim}).

The advantage offered by adaptive detection is only a constant factor 
(of $1/\sqrt{2}$) because so far I have discussed only coherent light. 
To break the SQL scaling $\sigma^{2} 
\sim {N}^{-1/2}$ it is necessary to use nonclassical light. The most 
obvious sort of nonclassical light to consider for this situation is 
broad-band squeezed light \cite{Bac98}. This is light where the  
spectrum of the dyne photocurrent of one quadrature, 
normalised as in \erf{dynepc}, has 
a noise spectrum $S$ below unity over some bandwidth broad compared 
to $\sqrt{\kappa\alpha}$.  
For moderate {\em phase quadrature squeezing} of spectral noise ${S}$, we 
can use the same feedback algorithm (\ref{genfb}) 
to get an improvement in the MSE  
by a factor of $\sqrt{S}$:
\beq
\sigma^{2} = \sqrt{S}/2\sqrt{N}~,~~{\rm for}~{
\chi}={\kappa}/\sigma^{2}.
\eeq

It would appear from this that the MSE in the phase estimate could be 
made as small as one desires by making the phase quadrature more 
squeezed. However this is not the case because the noise gets ``squeezed''
into the amplitude quadrature of the light \cite{Bac98,WalMil94}. Because the local 
oscillator phase $\Phi$ is only orthogonal to the {\em estimated} 
system phase $\hat\varphi$, not the true phase $\varphi$, the 
amplitude fluctuations {\em do} contribute to the noise in the phase measurement. 
It can be shown that as a consequence  there is an {\em optimal} 
non-zero squeezing spectrum ${S} \sim 
{N}^{-1/3}$. This is, in a sense, a ``more quantum'' feature than the 
existence of shot noise which gave the above limit (\ref{adaptSQL}), 
because it is a consequence of the uncertainty relation between 
amplitude and phase for a light beam \cite{WalMil94}. In any case, 
with squeezed light we can beat the SQL, and find a new quantum limit 
for CW phase locking:
\beq
 \sigma^{2} \sim {N}^{-2/3}.
\eeq

\begin{figure}
\includegraphics[width=0.47\textwidth]{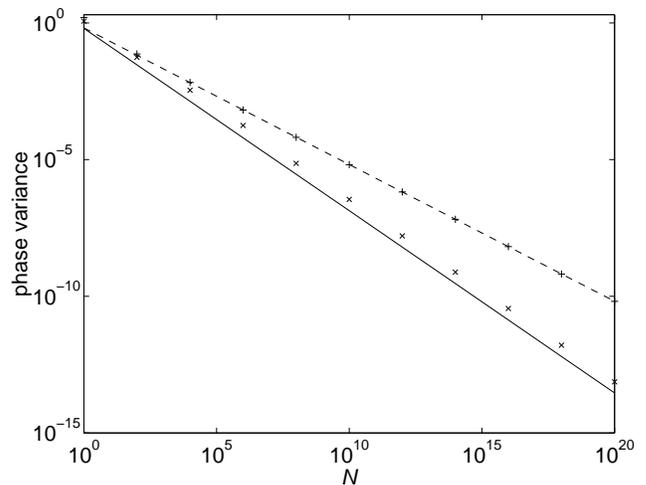}
\caption{Numerical results for the mean square phase error using optimal
squeezing, for heterodyne (+) and adaptive (x) detection. Note that these
were obtained using a more sophisticated feedback algorithm than that
explained above. The analytical asymptotic results are: heterodyne (-- --),
$0.66/N^{1/2}$ and adaptive (---), $0.63/N^{2/3}$.}
\end{figure}

\section{Experimental Progress}

Performing a quantum-limited adaptive phase measurement has been one 
of the primary research objectives of Assoc. Prof. Hideo Mabuchi since 
his appointment at 
CalTech in 1999. As mentioned in the introduction, this is the first step 
towards a longer-term goal of investigating quantum-limited control in general, 
and in particular in cavity QED systems. This first step has proven 
quite challenging in its own right!

Rather than a CW measurement of a varying phase as discussed above, 
Mabuchi's group has been working on single-shot estimates of an 
unknown phase imprinted on a pulse of light, as originally considered 
in Ref.~\cite{Wis95}. Here the parameter $\bar{n}$ (see table 1) is 
simply the mean number of photons in a single pulse on which a single 
unknown phase is imprinted. For an adaptive measurement, the phase 
must be estimated when only a fraction of these $\bar{n}$ photons have 
entered the detector. The fact that this can work even for $\bar{n}$ 
of order unity is described by Mabuchi as a ``pretty wild fact that 
makes you think twice about field quantisation''.

In this single-shot case the feedback algorithm is 
considerably more complicated than that in Sec.~IV, as the local oscillator phase 
$\Phi(t)$ may depend nonlinearly upon various integrals of the 
photocurrent {\em during the course of a single pulse} \cite{BerWis01a}. 
The form of these integrals can only be derived from the full quantum 
theory \cite{WisKil98}. Mabuchi  
thus decided to use digital electronics, and to discretise the feedback 
algorithm.

Just as the CW 
result requires the magnitude of the feedback $\chi$ to be close to 
its theoretical optimum (as 
discussed in Sec.~IV),  
the single shot case is sensitive to the design of the feedback loop. Through 
numerical modeling, Mabuchi's group have found that the discretisation requires 
at least of order 100 divisions of the total pulse length $T$. 

The 
only electronics fast enough to do the required processing are field 
programmable gate arrays \cite{Sto02}. 
These are so fast (10s of MHz) that the feedback 
bandwidth, of around 5 MHz, is actually limited by the the bandwidth of 
the electro-optics (in particular 
the synthesiser used to modulate the feedback). This bandwidth allows each 
time-division to be of order $500$ns, for a total time $T$ of only about 
$50\mu$s. 
This still required the production of a coherent state of this 
duration. In other words, it required light with a shot-noise limited 
amplitude spectrum down to about 50 kHz. This was another enormous technical 
challenge, as this is far below the usual frequency regime of 
quantum-limited experiments \cite{Bac98}. It was achieved using a very 
high Q cavity of lifetime 16 $\mu$s as a mode cleaner.

Because they are not (as yet) using any non-classical source, the 
experimental signature Mabuchi's group is seeking is a reduction of 
the MSE (by at most a factor of 2 --- see table 1) below the 
heterodyne limit of $0.5/\bar{n}$. In work just released to the 
physics archive \cite{Arm02}, this reduction has been convincingly 
demonstrated. Of course, much work 
remains to be done. For example, Mabuchi plans to investigate the
 tails of the distribution as well as its 
standard error, to verify some of the predictions of 
Ref.~\cite{WisKil98}.  Employing non-classical light to break the SQL 
is another longer term goal. Other experimental groups are interested 
in this as well.

\section{Conclusions}

Estimating an {\em unknown} phase ${\varphi}$ at the quantum limit 
is a difficult problem. There are many different cases that can be 
considered, but 
for all of them the standard quantum limits (SQLs) for the mean-square error 
arise for coherent light or non-adaptive measurements. 
The SQLs can be beaten only by using  nonclassical light 
{\em in conjunction with 
adaptive measurements}. When devices including phase-locking loops 
are sufficiently miniaturised, quantum limits will become practical 
limits. Thus, as well as being of theoretical 
interest, adaptive measurements should   become 
part of the tool-kit of the future ``quantum engineer'' who seeks to manipulate
quantum systems as well as nature allows.  

The improvement offered by adaptive detection over non-adaptive 
detection in phase estimation has very recently been achieved by 
the group of Hideo Mabuchi at CalTech. This experiment breaks new 
ground in the use of low-frequency optical mode cleaning and 
high-speed digital electronics in quantum-limited 
experiments. The theoretical and experimental 
techniques developed also pave the way for future 
research into the feedback-control of quantum optical systems in 
general.


\begin{thebibliography}{99}

\bibitem{Wis95}  
H.M. Wiseman, 
Phys. Rev. Lett. {\bf 75}, 4587 (1995).

\bibitem{WisKil97} 	H.M. Wiseman and R.B. Killip,
	Phys. Rev. A {\bf 56}, 944 (1997).
	
\bibitem{WisKil98}	H.M. Wiseman and R.B. Killip,
	Phys. Rev. A {\bf 57}, 2169 (1998).
	
\bibitem{BerWisZha99}
 	D. Berry, H.M. Wiseman, and Zhong-Xi Zhang,
    Phys. Rev. A {\bf 60}, 2458 (1999).
    
\bibitem{BerWis00}
D.W. Berry and H.M. Wiseman,
Phys. Rev. Lett. {\bf 85}, 5098 (2000).

\bibitem{BerWis01a} 
D. Berry and H.M. Wiseman,
Phys. Rev. A {\bf 63}, 013813 (2001).

\bibitem{BerWis01b} 
D.W. Berry and H.M. Wiseman,
J. Mod. Optics. {\bf 48},  797 (2001).

\bibitem{BerWisBre01}
D. W. Berry, H. M. Wiseman and J. K. Breslin,
Phys. Rev. A. {\bf 63}, 053804 (2001).

\bibitem{BerWis02}
D.W. Berry and H. M. Wiseman, 
Phys. Rev. A. {\bf 65}, 043803 (2002).

\bibitem{WalMil94}
D. F. Walls and G. J. Milburn,
{\em Quantum Optics}
(Springer, Berlin, 1994).

\bibitem{PegBar97}
D. T. Pegg and S. M. Barnett, 
J. Mod. Opt. {\bf 44}, 225 (1997).

\bibitem{HarSan96}
P. Hariharan and B. C. Sanders, 
Progress in Optics {\bf XXXVI}, 49 (1996).

\bibitem{Sto02}
J. Stockton, M. Armen, and H. Mabuchi,
``Programmable Logic Devices in Experimental Quantum Optics'',
quant-ph/0203143

\bibitem{Bac98}
H-A. Bachor, 
{\em A Guide to Experiments in Quantum Optics} 
(Wiley, 1998)

\bibitem{Arm02}
M. A. Armen, J. K. Au, J. K. Stockton, A. C. Doherty, and H. Mabuchi,
``Adaptive homodyne measurement of optical phase'', 
quant-ph/0204005
    
\end{thebibliography}
\end{document}